\documentclass[sigconf]{acmart}
\AtBeginDocument{%
  }

\setcopyright{acmlicensed}
\copyrightyear{2018}
\acmYear{2018}
\acmDOI{XXXXXXX.XXXXXXX}
\acmConference[EASE 2025]{The 29th International Conference on Evaluation and Assessment in Software Engineering}{17–20 June, 2025}{Istanbul, Türkiye}
\acmISBN{978-1-4503-XXXX-X/2018/06}

\setcopyright{none}
\renewcommand\footnotetextcopyrightpermission[1]{}
\usepackage{multirow}

\begin{document}

\title{Integrating Human Feedback into a Reinforcement Learning-Based Framework for Adaptive User Interfaces}

\author{Daniel Gaspar-Figueiredo}
\affiliation{%
  \institution{Universitat Politècnica de València and Instituto Tecnológico de Informática}
  \city{Valencia}
  \country{Spain}}
\email{dagasfi@epsa.upv.es}

\author{Marta Fernández-Diego}
\affiliation{%
  \institution{Universitat Politècnica de València}
  \city{Valencia}
  \country{Spain}}
\email{marferdi@omp.upv.es}

\author{Silvia Abrahão}
\affiliation{%
  \institution{Universitat Politècnica de València}
  \city{Valencia}
  \country{Spain}}
\email{sabrahao@dsic.upv.es}

\author{Emilio Insfran}
\affiliation{%
  \institution{Universitat Politècnica de València}
  \city{Valencia}
  \country{Spain}}
\email{einsfran@dsic.upv.es}

\renewcommand{\shortauthors}{Gaspar-Figueiredo et al.}

\begin{abstract}

Adaptive User Interfaces (AUI) play a crucial role in modern software applications by dynamically adjusting interface elements to accommodate users’ diverse and evolving needs. However, existing adaptation strategies often lack real-time responsiveness.
Reinforcement Learning (RL) has emerged as a promising approach for addressing complex, sequential adaptation challenges, enabling adaptive systems to learn optimal policies based on previous adaptation experiences. Although RL has been applied to AUIs,integrating RL agents effectively within user interactions remains a challenge.

In this paper, we enhance a RL-based Adaptive User Interface adaption framework by incorporating personalized human feedback directly into the leaning process. Unlike prior approaches that rely on a single pre-trained RL model, our approach trains a unique RL agent for each user, allowing individuals to actively shape their personal RL agent's policy, potentially leading to more personalized and responsive UI adaptations. 
To evaluate this approach, we conducted an empirical study to assess the impact of integrating human feedback into the RL-based Adaptive User Interface adaption framework and its effect on User Experience (UX). The study involved 33 participants interacting with AUIs incorporating human feedback and non-adaptive user interfaces in two domains: an e-learning platform and a trip-planning application. 
The results suggest that incorporating human feedback into RL-driven adaptations significantly enhances UX, offering promising directions for advancing adaptive capabilities and user-centered design in AUIs.

\end{abstract}



\maketitle

\section{Introduction}

Adaptive systems and Adaptive User Interfaces (AUI) address usability challenges by dynamically adjusting interface elements~\cite{Akiki:2014, Viano:2000}. Existing adaptation approaches, including recommendation systems and configurable UIs~\cite{jorritsma2015adaptive, Wei:ChronicDis:2024, langerak:marlui:2024}, often struggle with complex, long-term adaptation and holistic interface optimization.
Most existing research has focused on adapting isolated UI elements, such as menus~\cite{Todi:2021MCTS, Vanderdonckt:2016, Brumby:MenuArrange:CHI15}, rather than holistically optimizing the entire interface. This limitation highlights the need for approaches capable of handling the inherent complexity of adapting the entire interface dynamically. Furthermore, many existing adaptation strategies rely on pre-defined rules or offline models, which lack the flexibility to support real-time user interactions.

    Recent advancements in Artificial Intelligence (AI), particularly Machine Learning (ML), offer promising avenues to support more effective UI adaptation. Among these, Reinforcement Learning (RL) has emerged as a robust solution for managing the complexities of adaptive interfaces. Unlike traditional ML approaches that optimize specific UI features using static adaptation models (or do not learn from past actions over time), RL enables an agent to learn adaptive policies that evolve over time by continuosly refining its decisions based on past interactions, which can be useful to guide changes across the entire interface. By modeling the UI adaptation problem as a sequential decision-making process, an RL agent can optimize key quality attributes, such as User Experience (UX), over extended interaction periods, by maximizing cumulative rewards that reflect overall user experience.
    
    

However, designing effective RL-based UI adaptations requires careful consideration of the reward function, which quantifies the success of each adaptation and guides the agent toward beneficial adjustments. A poorly designed reward function can lead to suboptimal behaviors, where the agent prioritizes factors that do not align with user needs. To address this problem, a new and growing area of RL research explores the integration of \textit{human feedback (HF)}, where humans directly influence the agent’s learning process by providing input on its decisions. Human feedback offers a mechanism to refine the reward signal, ensuring that the agent's learned adaptations align more closely with user preferences. Despite its potential, effectively incorporating human feedback into RL-based adaptation remains an open challenge.

Building upon the RL-based UI adaptation framework proposed by Gaspar-Figueiredo et al.~\cite{gasparEICS:2024}, we aim to address its limitations by incorporating human feedback directly into the learning and adaptation process. The original framework lacked mechanisms for capturing user preferences and was solely evaluated in simulated environments, limiting its applicability to real-world interactions. Our initial extension~\cite{gasparEMSE:2024} introduced human feedback but applied a single HF model across all users, failing to account for individual preferences and resulting in suboptimal adaptation outcomes. 


Neglecting individualized human feedback in adaptive systems can result in user interfaces that fail to respond effectively to actual user preferences. Without direct user input, RL agents must rely on generalized data or simulated feedback, which often fails to capture individual behaviors and contextual requirements. This limitation reduces the system’s ability to fine-tune adaptations based on real-world interactions, potentially missing opportunities for deeper personalization~\cite{bach2024systematic}. Addressing this gap requires leveraging human feedback mechanisms to refine RL agents at an individual level, ensuring that adaptive UI strategies align more closely with user preferences and ultimately improve user satisfaction..

In this paper, we therefore enhance the RL-based framework for UI adaptations proposed by Gaspar-Figueiredo et al.~\cite{gasparEICS:2024} by integrating human feedback directly into the learning and adaptation process, enabling users to shape adaptations in real-time. Additionally, to evaluate the effectiveness of this approach, we report the results of an empirical study aimed at evaluating the impact of Human Feedback on RL-based UI adaptations and their effect on user experience, using non-adaptive interfaces as a baseline for comparison. Since non-adaptive systems provide a fixed and consistent experience, they serve as a suitable benchmark for assessing the effectiveness of RL-based AUIs. 


   This paper is organized as follows: Section 2 review existing empirical studies on adaptive user interfaces. Section 3 provides an overview of the original RL-based UI adaptation framework and details our extension to incorporate human feedback at an individual level (by each user). Section 4 describes the design and execution of the empirical study, while Section 5 presents the results and explores their implications for research and practice. Section 6 addresses potential threats to the validity of the findings. Finally, Section 7 concludes the paper and outlines future research directions.

\section{Related Work}

The Software Engineering (SE) and Human-Computer Interaction (HCI) communities communities have made significant progress in advancing the design and evolution of AUIs by leveraging AI techniques.  Despite positive results in some studies~\cite{Zouhaier:AUISaccessibility:2023}, challenges remain, including unpredictability, detrimental adaptations, inaccurate interpretation of user needs, and limited user involvement. A fundamental aspect of applying RL to AUIs is defining an appropriate reward function, which is challenging, especially when user goals are ambiguous or the impact of actions is unclear.

Prior research has explored various RL approaches for UI adaptation. Langerak et al.~\cite{langerak:marlui:2024,langerak:2022marlui}proposed a multi-agent RL framework. Vidmanov and Alfimtsev [35] integrated a usability reward model. Todi et al. [32] employed Monte Carlo Tree Search (MCTS) with predictive HCI models.  The authors in [16] integrated predictive HCI models into a RL-based approach, demonstrating RL's potential to automate UI adaptation based on usability metrics.  However, this study relied on simulations, failing to capture the complexities and variability of real-world user behavior.


Prior research has explored various RL approaches for UI adaptation.
Langerak et al.~\cite{langerak:marlui:2024,langerak:2022marlui} proposed a multi-agent RL framework in which a user agent learns to interact with the UI while an interface agent learns to adapt it. Vidmanov and Alfimtsev~\cite{Vidmanov:2024} extended this approach by integrating a usability reward model, enabling cooperative adaptation of UI elements to enhance user experience. Similarly, Todi et al.~\cite{Todi:2021MCTS} employed Monte Carlo Tree Search (MCTS) combined with predictive HCI models to simulate and evaluate UI adaptation strategies. By leveraging a pretrained value network, their approach mitigated the computational cost of online simulations while maintaining a high success rate.


Similarly, in~\cite{gasparEICS:2024}, the authors integrated predictive HCI models into a RL-based approach for AUIs. This study trained and evaluated RL agents in a simulated environment to assess their effectiveness in adapting UIs, specifically aiming to maximize user engagement.
The results demonstrated RL's potential to automate UI adaptation based on general-purpose usability metrics. However, a major limitation of this study was its reliance on simulations, without involving user interactions. Consequently, the study was unable to capture the complexities and variability of user behavior that occur in real-world scenarios, where individual preferences and interactions can vary widely.

To the best of our knowledge, no prior work has applied Reinforcement Learning with Human Feedback (RLHF) to AUIs. Building on this foundation, we extended the framework~\cite{gasparEICS:2024} by incorporating human feedback, distinguishing our approach from prior work by explicitly integrating user preferences into the adaptation process. While previous studies have explored RL-based UI adaptation through predictive models and automated agents, our work emphasizes direct human feedback to refine adaptation strategies, thereby aligning the adaptations more closely with user needs. 

In a previous empirical study~\cite{gasparEMSE:2024}, we used a general model, trained solely on feedback provided by the experimenter (when training the RL models), to incorporate human feedback into the adaptation process. This approach aimed to assess whether user feedback, even provided only by the experimenters, could enhance user satisfaction and user engagement when interacting with AUIs. Specifically, the study compared AUIs which used predictive HCI models and AUIs which used predictive HCI models augmented with human feedback to non-adaptive interfaces. 
The findings suggested a potential for human feedback to improve user satisfaction and engagement, despite lacking statistical significance. However, the study had limitations, primarily due to the use of a general feedback model that may have been insufficiently sensitive to individual user preferences. The experimenter’s preferences, which shaped the model, may not have accurately reflected the diversity of the participant base.


This paper addresses the limitations of using a general preference model by incorporating personalized RL agents trained by each user. 
Unlike the previous study~\cite{gasparEMSE:2024}, where a general model was trained based on experimenter-provided feedback, we now introduce models that learn directly from individual users, allowing the system to adapt more precisely to their unique preferences and needs. By comparing AUIs with personalized feedback models to non-adaptive interfaces, this study aims to deepen our understanding of how user-specific adaptations impact user satisfaction and engagement. This progression illustrates the evolution of the framework—from initial simulations, to general feedback, and now to a fully personalized, “human-in-the-loop” adaptation strategy.

\section{RL-Based UI Adaptation Framework}
\label{sec:framework}

\begin{figure*}
    \centering
    \includegraphics[width=0.8\linewidth]{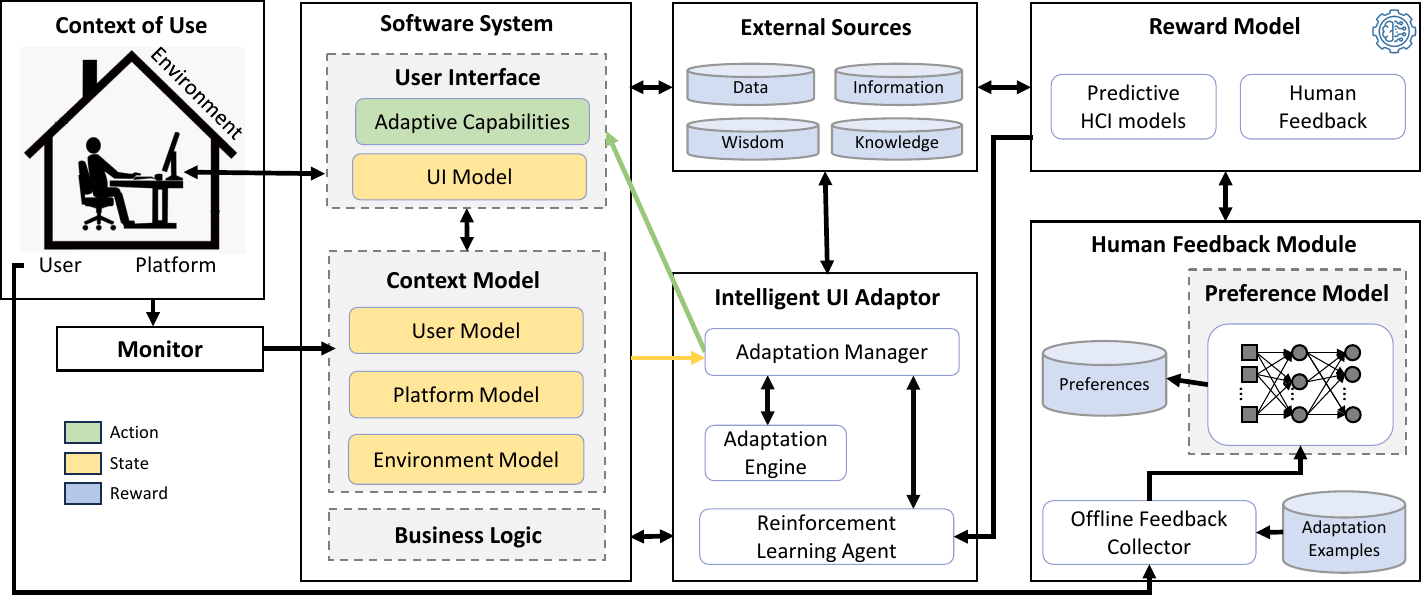}
    \caption{Extended intelligent UI adaptation framework from~\cite{abrahaoModel:2021}. Human Feedback Module and Reward Model are the main contributions in this paper.}
    \label{fig:framework}
      \Description{Intelligent User Interface adaptation framework. This figure shows how multiple components such as Software system, Intelligent UI adaptor, External Sources, Reward Models, Human Feedback Module and Context of use interact in order to achieve interfaces that learn to adapt based on Human Feedback.}
\end{figure*}

This section presents the RL-based UI adaptation framework. First, we provide an overview of the original framework~\cite{abrahaoModel:2021}, detailing its components and their functions. Subsequently, we explain how we extended this framework to incorporate human feedback directly into the learning process.


\subsection{Overview of the Original Framework}

As shown in Figure~\ref{fig:framework}, the framework consists of four main parts: \textit{Context of Use}, \textit{Software System}, \textit{Intelligent UI Adaptor}, and \textit{External Sources}.

The \textit{Context of Use} represents the user(s) interacting with their platform or device in any physical environment~\cite{calvary:2003}. The \textit{Monitor} component senses and abstracts relevant data into model fragments (\textit{Context Model}) within the \textit{Software System}.
The \textit{user model} captures end-user data (e.g., gender, age, interaction history, abilities, preferences, emotional state). The \textit{platform model} captures platform data (e.g., screen resolution, sizes, interaction capabilities, CPU availability). The \textit{environment model} captures environmental data influencing the UI (e.g. location, light, noise level)~\cite{abrahaoModel:2021}. 

Within this architecture, the \textit{Intelligent UI Adaptor} includes key elements that drive the adaptation process: the \textit{Adaptation Manager}, \textit{Adaptation Engine} and \textit{Adaptation Machine Learning} (subsequently formalized as a \textit{Reinforcement Learning Agent} in~\cite{gasparEICS:2024}).
This agent operates within a Markov Decision Process (MDP), optimizing adaptations based on \textit{state} information (drawn from the \textit{UI Model} and \textit{Context Model}), \textit{actions} (defined by \textit{Adaptation Capabilities}), and \textit{reward} (sourced from \textit{External Sources}).

The \textit{Adaptation Manager} orchestrates the adaptation process, managing initiation, execution, and completion to align adaptations with user needs. The manager works with the \textit{RL agent} to execute adaptations based on learned policies, using the adaptation logic in the \textit{Adaptation Engine}, which is usually implemented in the form of adaptation rules. The \textit{Adaptation Manager} also monitors adaptation outcomes, providing feedback to the \textit{RL agent} to refine its strategies. This real-time feedback loop allows the \textit{RL agent} to evaluate the success of its decisions and refine its strategies, ensuring that the system learns which adaptations are most effective and preferred by the user.



\subsection{Extending the RL-based Framework for AUI}

We extended the original framework by introducing new components and mechanisms to support more personalized, human-centered adaptations. Key contributions include enhancements to AUIs, the inclusion of a \textit{Human Feedback Module} to predict adaptation rewards through \textit{Reward Modelling}, and its implementation.

\subsubsection{\textbf{Adaptive User Interfaces}}

To illustrate the extended framework's flexibility, consider an example implementation of adaptive user interfaces across different application domains\footnote{Available at \url{https://github.com/RESQUELAB/Adaptive-app}}. As part of the \textit{Software System}, the user interface must include the \textit{Adaptation Capabilities}. These adaptations are guided by a \textit{RL agent} trained to optimize user experience.

For instance, the UI adaptive capabilities include:

\begin{itemize}
    \item \textit{Layout Adjustments}: Reorganizing elements for readability or emphasis, such as vertical lists or grid layouts (e.g., two to five columns).
    \item \textit{Font Size Modifications}: Adjustments for readability, benefiting users with visual impairments.
    \item \textit{Content Density Control}: Adjusting the amount of information displayed, from detailed content to condensed summaries.
    \item \textit{Theme Adjustments}: Modifying the color scheme and visual theme to match user preferences and reduce eye strain.
    \item \textit{Widget Customization}: Adapting navigation menus between list and dropdown formats.
\end{itemize}






By standardizing these adaptation actions across domains, the framework supports consistent evaluation and comparison of their impact on UX metrics. It is important to note that these represent examples of adaptations implemented for this study; however, the framework is not constrained to these specific actions. Its design allows for the incorporation of additional or alternative adaptations for other domains or application requirements.

\subsubsection{\textbf{Reward Models and Human Feedback for AUIs}}

We extended the framework by integrating a dual-source \textbf{reward}~\cite{christiano:deeprlhf:2023} mechanism for the \textit{RL Agent}, aligning the adaptation process with both general knowledge and user preferences. As shown in Figure~\ref{fig:framework}, the \textit{Reward Model} uses two types of sources. The first reward source uses baseline \textit{predictive Human Computer Interaction models}, developed using ML algorithms applied to data, information, wisdom, and knowledge from \textit{External Sources}, to provide a belief about user status. The second reward source incorporates \textit{Human Feedback} from the \textit{Human Feedback Module}, serving as a reward modifier that directly influences adaptation outcomes to reflect user satisfaction and preferences. This feedback allows the RL Agent to refine its strategies.


Human feedback is obtained through the \textit{Offline Feedback Collector}, which supports multiple standardized feedback encoding formats, including comparative feedback, attribute feedback, evaluative feedback, visual feedback, and keypoint feedback (see~\cite{yuan:rlhfTypes:2024} for detailed descriptions). To elicit this feedback, users are presented with a curated set of \textit{Adaptation Examples} generated in advance, rather than during live system interaction. This design supports an offline learning process, where user preferences are gathered in batch and subsequently processed by a \textit{Preference Model}. The model can be trained using various neural network architectures, such as Multilayer Perceptrons or Convolutional Neural Networks~\cite{yuan:rlhfTypes:2024}.

Each user that provides feedback will have their own personalized \textit{Preference Model}, which evolves dynamically as more feedback is collected. This model ensures that the adaptation process becomes increasingly customized to the individual's preferences over time, leading to a system capable of fine-tuning its decisions to align with the specific needs and expectations of each user.

By embedding user feedback directly into the reward structure, the framework adopts a "human-in-the-loop" approach, transforming it into a cooperative interaction loop between the user and the system. This integration allows the adaptive system to move beyond a generic adaptation strategy, creating a cooperative interaction loop where the agent continuously refines its decisions based on specific user input. By coupling baseline HCI metrics with human feedback, 
we aim to address the gap in intelligent UI adaptation: \textit{the need for interfaces that are both technically efficient and meaningfully aligned with the human aspects of interaction}~\cite{abrahaoModel:2021}.

\subsubsection{\textbf{Implementation}}

We used the replication package provided by Gaspar-Figueiredo et al.~\cite{gasparEICS:2024}, which utilizes OpenAI Gym~\cite{openaigym:github:2016} to offer a structured environment for RL implementations. Using OpenAI Gym, we defined the system's states (\textit{Context Model} and \textit{UI Model}), actions (\textit{Adaptation Capabilities}), and reward mechanisms. These elements ensure compatibility with RL algorithms such as Q-Learning~\cite{gasparEICS:2024} and Monte Carlo Tree Search~\cite{browne:mcts:2012}. The extended framework is available here: \url{https://github.com/RESQUELAB/RL-Based-Framework}

\section{Experimental Study}
\label{sec:experimental_study}

Following the GQM template for goal definition~\cite{Basili:1994:GQM}, the goal of this study is to \textbf{analyze} the integration of Human Feedback from individual users into a RL-based UI adaptation approach \textbf{with the purpose of} assessing its impact as regards adapting UIs \textbf{with respect to} its ability to improve the user experience \textbf{from the point-of-view of} both novice developers and researchers interested in adaptive user interfaces.
    

The study involves Master’s students in Computer Science at the \textit{Universitat Politècnica de València (UPV)}, interacting with UIs from two domains: an \textit{e-learning system} and a \textit{trip-planning system}. These domains test UI adaptation and show how domain specificity influences user feedback and adaptations.
    

\subsection{Research Question and Hypothesis}

    We aim to investigate the extent to which incorporating human feedback into RL-based UI adaptations affects the user experience across different domains. User experience is operationalized in terms of user satisfaction and user engagement (see the next subsection for details). The research question (RQ) and hypotheses guiding the investigation are as follows:
    
     $\textbf{RQ1}$: How does incorporating human feedback into a reinforcement learning-based UI adaptation affect the overall user experience?

    \begin{itemize}
        \item $H_{n1}$: There is no significant difference in user satisfaction between AUIs that use predictive HCI models with personalized human feedback and non-adaptive UIs.
        \item $H_{n2}$: There is no significant difference in user engagement between AUIs that use predictive HCI models with personalized human feedback and non-adaptive UIs.
    \end{itemize}

The statistical analysis aims to reject these hypotheses and accept the alternative ones (\textit{e.g.,} $H_{a1}$ = $\neg H_{n1}$). These two-sided hypotheses do not postulate any effect from different adaptation strategies.


Non-adaptive interfaces serve as a baseline to evaluate the impact of RL-based UI adaptations, isolating the contributions of RL mechanisms and human feedback. This approach also reflects real-world applications where static designs are common. Prior work showed no statistically significant difference among Adaptive and non-adaptive approaches, highlighting the need for further research on how RL-based adaptations and human feedback influence user experience.

       
\subsection{Variables}

This section outlines the variables, measurement scales, and their operationalization.

    
\subsubsection{\textbf{Independent Variables}}

The main independent variable is the \textit{UI adaptation strategy}, a nominal variable with two levels: Adaptive UI with human feedback (Adaptive) and Non-Adaptive UI (NA).
%
%
The secondary independent variable is the \textit{application domain}, a nominal variable with two levels: \textit{Trips} (a travel planning application) and \textit{Courses} (an e-learning system). 

\subsubsection{\textbf{Dependent Variables}}
\label{sec:dependentVar}

The primary dependent variable is UX, defined by ISO 9241-210: 2010 as \textit{"a person’s perceptions and responses that result from the use and/or anticipated use of a product, system, or service."} Unlike task-oriented interactions, UX encompasses a broader focus on “experience,” which includes hedonic qualities, emotions, and effects (e.g., feelings of interest, enthusiasm, or irritation) that users encounter while interacting with software systems. Given its multifaceted nature, UX can be operationalized in various ways. In this study, we assessed UX through two key dimensions: user satisfaction and user engagement.


    \textit{User satisfaction} is a measure of the quality of UX and reflects the extent to which a user’s expectations are met~\cite{iso9241:2018}. It was assessed using the \textit{Questionnaire for User Interaction Satisfaction (QUIS)}~\cite{norman:1998}, evaluating user acceptance of a computer interface across factors like ease of use, consistency, system capability, and learnability, rated on a 10-point scale, and an aggregated score was calculated for each participant.

    \textit{User engagement} reflects the quality of user interaction with software systems, including aesthetics, sensory appeal, perceived control, time awareness, motivation, and emotional responses~\cite{obrien:2008, doherty:2018, LEHMANN:2012}. It was measured using the \textit{User Engagement Scale (UES) questionnaire} \cite{Obrien:UES:2018}, a 31-item questionnaire assessing six dimensions: focused attention, perceived usability, aesthetic appeal, endurability, novelty, and involvement. Each item was rated on a five-point Likert scale, and an aggregated user engagement score was calculated for each participant.
    

\subsection{Subjects}

Participants were recruited using convenience sampling, a non-probability method selecting individuals based on availability~\cite{Baltes:2022}. The study targeted Master’s students in computer science 
at the \textit{Universitat Politècnica de València}.


A total of $33$ participants ($23$ male, $7$ female, and $3$ Non-Binary) were recruited. 15 participants work as software developers. Participation was voluntary, and participants received informed consent detailing the study's purpose and procedures.
    %
    %
    %
    The study was conducted on-site and adhered to the 
    \textit{UPV's}
    IRB protocol. The consent form explicitly stated that no personal data would be collected; if any data were to be collected, it would not be published and would be destroyed after the study's conclusion.

\subsection{Design}

\begin{figure*}
    \centering
    \includegraphics[width=0.89\linewidth]{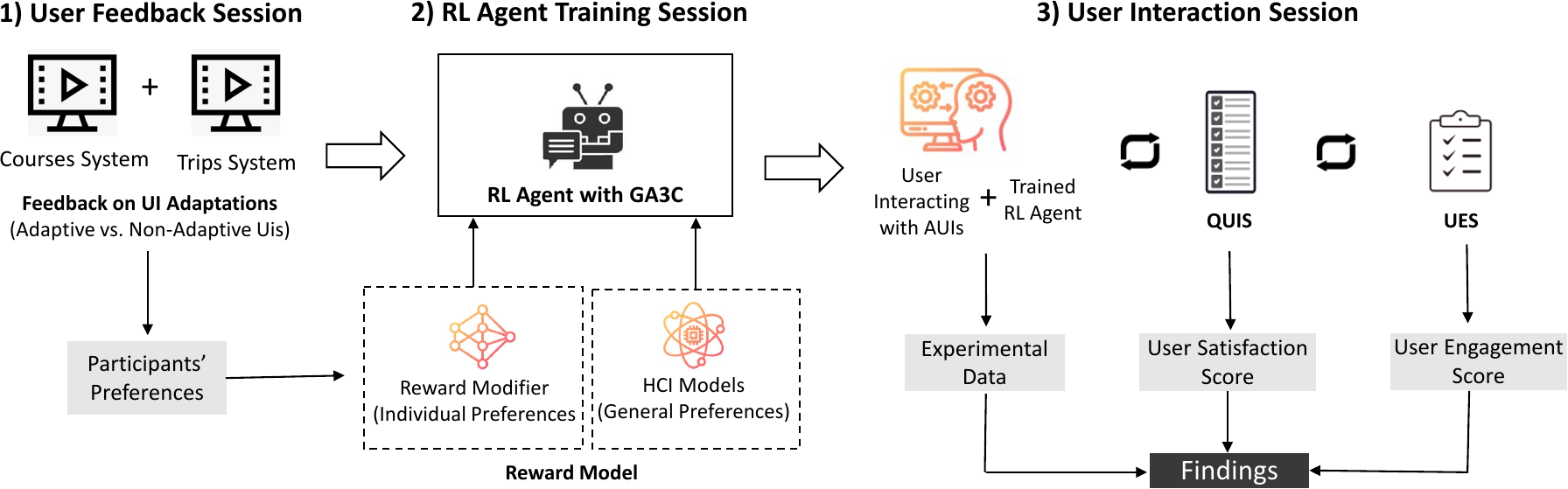}
    \caption{Experimental methodology: 1) Feedback collection; 2) Individualized RL agent training; and 3) User interaction with AUIs and non-adaptive UIs}
    \label{fig:researchMethodology}
    \Description{The experimental methodology starts with a User Feedback Session, where users provide feedback on 2 different domains (Courses and Trips). This feedback produces the Participants' preferences, which feeds a Reward Modifier. On the second step of this methodology, this modifier, with HCI models feed the learning of a RL agent with GA3C. Finally, with a trained RL agent for each user, the users interact with the adaptive systems and provide feedback with QUIS and UES questionnaires, which will provide the findings of the experiment.}
\end{figure*}

Figure~\ref{fig:researchMethodology} provides an overview of our study methodology and outcomes, which unfold across three sessions. In Session 1 (\textit{User Feedback}), participants’ preferences regarding AUIs are collected. Session 2 (\textit{RL Agent Training}) involves training a personalized RL agent for each participant, incorporating both predictions from an HCI model and individual preferences from Session 1. Finally, in Session 3 (\textit{User Interaction}), participants interact with both adaptive and non-adaptive UIs to evaluate their impact on UX. The following sections detail each session.


\subsubsection{\textbf{User Feedback Session}}
\label{sec:userFeedback}

To collect human feedback, we developed a web app\footnote{Available at \url{https://github.com/RESQUELAB/RL-Teacher-UIAdaptation}} that facilitates the display of video clips showcasing multiple adaptations in sequence. We chose comparative feedback for its simplicity and universality~\cite{yuan:rlhfTypes:2024}. In this implementation, video clips were presented for pairwise comparisons, allowing users to directly provide feedback on their preferences through relative comparisons. Each video depicted the evolution of a user interface through various adaptations, including changes in layout, font size, content density, theme, and widget type. These adaptations were shown across two domains: Courses and Trips. A total of 64 video clips (32 per domain) were generated by the experimenter and reviewed by another researcher with expertise in UI design. The participants evaluated multiple pairs of video clips, with the system using a red-black tree~\cite{Elmasry:redBlackTrees:2019} to organize the clips and iteratively requested feedback from participants to rank them according to their preferences. Participants continued evaluating pairs of clips until all clips were sorted and ranked. They assessed each adaptation based on usability and aesthetic appeal perceptions. Each clip lasted approximately four seconds to minimize fatigue and maintain  engagement.

\subsubsection {\textbf{RL Agent Training Session}}

RL agents were trained using the GA3C (GPU Asynchronous Advantage Actor-Critic) algorithm~\cite{babaeizadeh2016ga3c}, an extension of A3C (Asynchronous Advantage Actor-Critic) with GPU acceleration. GA3C handles high dimensional state and action spaces in UI adaptation by processing multiple asynchronous agents in parallel on GPU threads, incorporating user-specific preferences from feedback. The actor-critic framework refines the policy (actions taken) and value estimates (expected outcomes) based on participant input.


Personalized adaptation behavior was achieved using a \textit{dual reward structure}~\cite{christiano:deeprlhf:2023}. The primary reward was based on a predictive HCI model for general user interaction patterns. The reward modifier, generated by a Fully Connected Multi-Layer Perceptron trained on feedback from the video comparison session (Section \ref{sec:userFeedback}), adjusted the baseline reward to align the agent's behavior with user preferences, creating a unique RL agent for each user.


\subsubsection {\textbf{User Interaction Session}}

A balanced within-subject two-treatment factorial crossover design was used, where each participant interacted with adaptive and non-adaptive techniques across two domains (Courses and Trips). The order of techniques is randomized to control for ordering effects. Each session was followed by QUIS and UES questionnaires. This design addresses small sample sizes and increases the sensitivity of experiments~\cite{vegas:2015}.


    In a balanced crossover design, the measures are taken from a participant several times (i.e. a participant is assigned to a sequence of treatments). The crossover design was defined by following the guidelines proposed by Vegas et al. \cite{vegas:2015}. We specifically analyzed not only the effects of the treatments (i.e. Adaptive, NA) on the dependent variables, but also the effects of critical crossover variables (i.e. period, sequence and carryover). 

    Each group followed a different sequence of adaptation techniques and domains, producing four different sequences (see Table \ref{tab:crossover}).
    In the first period 1,
    participants in Group 1 began their interaction with the non-adaptive system in the Courses domain, while Group 4 also used the non-adaptive system but within the Trips domain. In contrast, in the same period, participants in Group 2 interacted with the adaptive app in the Trips domain, and Group 3 used the adaptive app in the Courses domain. Following these interactions, each of the four groups completed the QUIS and UES questionnaires to assess their levels of satisfaction and engagement with the respective adaptation technique and domain.
    
    The second period, maintained a similar structure, with Groups 2 and 3 switching to the non-adaptive systems, specifically using the Courses and Trips domains, respectively. Meanwhile, Groups 1 and 4 transitioned to the adaptive systems in the Trips and Courses domains, respectively.
    Similarly, after their interactions, the four groups completed the QUIS and UES questionnaires.

    These sequences were carefully selected to ensure that potential order effects would be minimized. 
    This crossover design makes it possible to detect carryover effects by analyzing the data separately for each order to see whether it had an effect on the results, thus ensuring that any potential bias due to the sequence of treatments can be identified and accounted for.


\begin{table}
\centering
\caption{Study design for session 3} 
\label{tab:crossover}
\begin{tabular}{|l|l|l|l|l|} 
\hline
\multirow{2}{*}{\textbf{Group}} & \multicolumn{2}{c|}{\textbf{Period 1}} & \multicolumn{2}{c|}{\textbf{Period 2}} \\ 
\cline{2-5}
                        & \textbf{Technique} & \textbf{Domain} & \textbf{Technique} & \textbf{Domain} \\ 
\hline
1                       & NA        & Courses             & Adaptive  & Trips                \\ 
\hline
2                       & Adaptive  & Trips               & NA        & Courses              \\ 
\hline
3                       & Adaptive  & Courses             & NA  & Trips                \\ 
\hline
4                       & NA        & Trips               & Adaptive        & Courses              \\
\hline

\multicolumn{5}{l}{\rule{0pt}{0.5em}\textit{Techniques: Adaptive / Non-Adaptive (NA)}}\\

\multicolumn{5}{l}{\rule{0pt}{0.5em}\textit{Domains: E-learning (Courses) / Trip-planning (Trips)}}\\

\end{tabular}
\end{table}

\subsection{Instrumentation}

The experimental setup included framework implementation (Section~\ref{sec:framework}), registration, consent agreements, and task instructions. The registration process assigned each participant to a personalized RL agent. Participants completed a demographic form and provided informed consent, outlining the study’s objectives, data protection, and ethical standards compliance. Participants received instructions for tasks in each application domain (Trips and Courses), including key interactions such as consulting personal information, browsing items, and making selections.
User experience was assessed using standardized questionnaires (Section~\ref{sec:dependentVar}).

\subsection{Execution}

The empirical study spanned two weeks, with two phases. Phase 1 involved preparatory tasks, including feedback collection and RL agent training. Phase 2 involved the user study, where participants interacted with \textit{Adaptive} and \textit{NA} interfaces across the \textit{Courses} and \textit{Trips} domains. Each participant attended two sessions in total, ensuring balanced exposure to the different interfaces and domains.


\subsubsection{\textbf{Feedback Session}}

The human feedback collection phase involved participants evaluating pre-generated UI adaptation clips from simulation environments.


Participants registered in a system linking them to an RL agent. Each participant reviewed $64$ UI adaptation clips ($32$ from each domain: \textit{Courses} and \textit{Trips}) and provided pairwise comparisons to indicate their preferred adaptation. This resulted in $8,356$ human feedback responses. Responses were balanced using a red-black tree structure, a type of self-balancing binary search tree~\cite{Elmasry:redBlackTrees:2019}, so the number of responses per participant varied.

    \subsubsection{\textbf{RL Agent Training Session}}

    The RL agent training phase was conducted on a high-performance computer equipped with an Intel Core i9-13900KF CPU, 64 GB DDR5 RAM, and an NVIDIA GeForce RTX 4090 GPU to handle the computational demands.

    Over three days, each participant’s RL agent underwent \textit{1,000,000} training steps using the GA3C algorithm. The training process employed a dual-reward system: a predictive HCI model trained on interaction data collected from 25 users interacting with an e-commerce interface under various UI configurations (e.g., grid vs. list layout, light vs. dark theme), used to model general engagement patterns~\cite{gasparEMSE:2024}; and a personalized reward modifier trained on individual feedback. This iterative approach ensured that each RL agent was uniquely tailored to reflect its assigned participant’s specific preferences.

    

\subsubsection{\textbf{User Interaction Session}}

Participants used their pre-assigned RL agent. Participants, already registered and assigned to a group, accessed the adaptive or non-adaptive systems in the sequence (domain and technique) based on their group.

    
    Each session consisted of the following steps:
        \begin{enumerate}
        
        \item Introduction and explanation of the task: The experimenter introduced the session and explained the task to the participants.
    
        \item Task execution: The participants performed the tasks using the UI and the technique assigned to them. The tasks to be performed included:
        
        \begin{itemize}

            \item Trip Planner system: Participants were asked to imagine planning a 5-day trip with a limited budget and explore the system for the best option.
            

            \item Course Management system: Participants were asked to imagine they wanted to take online courses to improve their skills in a specific area, with a limited budget to add as many courses as possible. The experimenter offered clarifications to avoid confusion.
                    
            
        \end{itemize}
        
        \item Post-task questionnaires: Upon completion of each session, participants filled out the QUIS and UES questionnaires through a dedicated page in the adaptive application, evaluating their satisfaction and engagement with each UI.    
    \end{enumerate}

Sessions 1 and 3 were conducted in a controlled lab environment, with uniform equipment and software configurations. Each participant was provided with a desk setup consisting of a computer screen, mouse, and keyboard. The order of the experimental sessions was counterbalanced to control for potential order effects. Each session lasted, a most, 120 minutes, though participants were given as much time as they needed to complete their tasks without time constraints.

\subsection{Analysis Plan}
\label{sec:analysisPlan}

QUIS and UES questionnaire responses were analyzed to assess participants' perceptions of engagement and satisfaction. Internal consistency was tested using Cronbach's alpha (threshold $\alpha \geq 0.70$)~\cite{Maxwell2002Applied}.


    We employed a combination of descriptive statistics, violin plots, and statistical tests to analyze the data collected during the experiment. Since the study utilized a crossover design, in addition to analyzing the primary experimental factors, it was essential to include period, sequence, and potential carryover effects. Hypotheses were tested using a Linear Mixed Model (LMM), following the guidelines established by Vegas et al.~\cite{vegas:2015}. 
    In the model, the following fixed factors were included: adaptation technique (treatment), domain (experimental object), and group or sequence (which was confounded with carryover and treatment*period interaction). Subject was included as a random factor nested within the sequence.
    
    The LMM allowed us to determine whether these factors significantly influenced the results. Assumptions for LMMs were thoroughly tested and reported. Specifically, we confirmed that the residuals of the models met the normality requirement by applying the Shapiro-Wilk test~\cite{vegas:2015}. Additionally, we used the Levene’s test for equality of variances to ensure homogeneity among residuals.

    The LMM was applied to each dependent variable in order to assess whether the technique, domain or sequence (confounded with carryover and treatment*period interaction) had statistical significance. This represents the probability that the result could have occurred by chance due to a Type I error.

\section{Results}

This section presents the quantitative analysis of the empirical study results. All data analyses were conducted in R version 4.4.0~\cite{team:2013r}. The linear mixed model analysis was run using the lmerTest package~\cite{lmerTest}, which provides p-values for tests for fixed effects.

\subsection{Descriptive Statistics and Exploratory Data Analysis}

Table~\ref{tab:technique_comparison} summarizes the descriptive statistics for \textit{User Satisfaction} and \textit{User Engagement}, grouped by the Adaptive UI with human feedback (Adaptive) and Non-Adaptive UI (NA) techniques. Both variables showed a higher mean for the adaptive interface, indicating a tendency toward increased satisfaction and engagement when using the adaptive UI compared to the non-adaptive UI. 
    
\begin{figure}[t]
    \centering
    \includegraphics[width=0.45\textwidth]{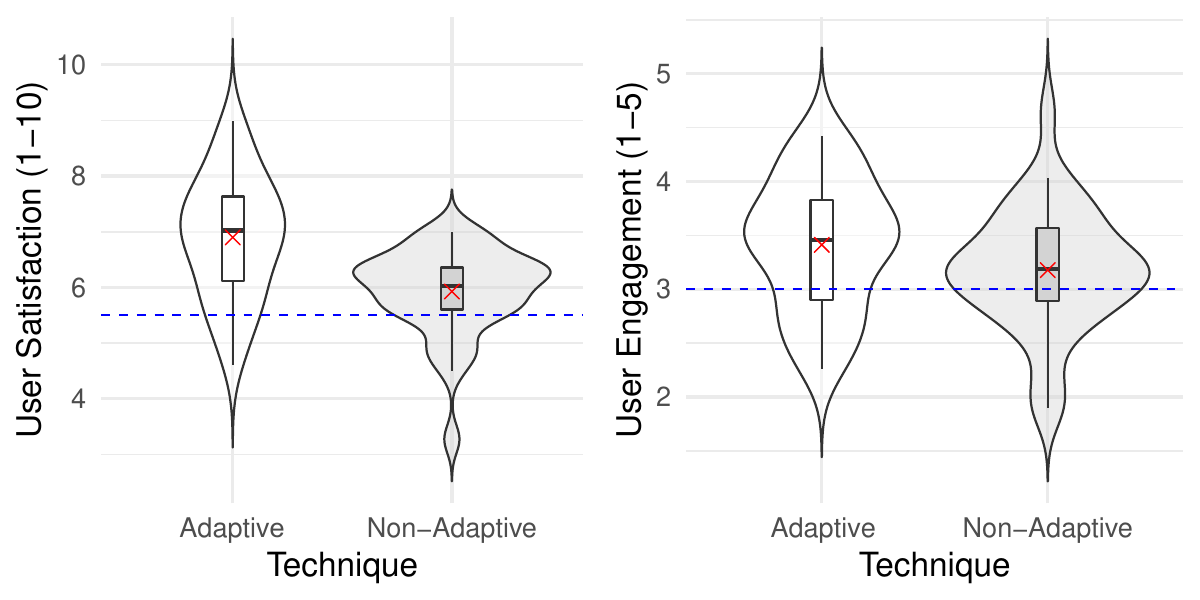}
    \caption{Violin plots of User Satisfaction and User Engagement by Technique. \textit{The blue dashed line represents the neutral midpoint for each scale. The red cross marker indicates the mean value for each technique.}}
    \label{fig:violin}
    \Description{Violin plots describing the distribution of User Satisfaction and User Engagement along the 2 techniques. There is a blue dashed line representing the neutral middle point.}
\end{figure}

    The distribution of scores is further illustrated through violin plots in Figure~\ref{fig:violin}, which provides a visual comparison of \textit{User Satisfaction} and \textit{User Engagement} scores between the two techniques. The \textit{User Satisfaction} scores, measured on a scale from 1 to 10, show distinct distributions between the two techniques, indicating differences in perceived satisfaction. Similarly, \textit{User Engagement} scores, assessed on a scale from 1 to 5, also demonstrate variation, with the \textit{Adaptive} technique appearing to elicit more consistent and higher levels of engagement compared to the Non-Adaptive approach.
    The blue dashed line in each plot represents the midpoint of the respective scales. Both techniques exceed this threshold suggesting that, regardless of the observed differences, users generally rated their satisfaction and engagement as favorable.

\begin{table}[ht]
\centering
\caption{Descriptive statistics for User Satisfaction and User Engagement by Technique}
\label{tab:technique_comparison}
\begin{tabular}{|p{1.5cm}|p{1.4cm}|p{0.5cm}|p{0.5cm}|p{0.7cm}|p{0.6cm}|p{0.5cm}|}
\hline
\textbf{Variable} & \textbf{Technique} & \textbf{Min} & \textbf{Max} & \textbf{Mean} & \textbf{Med} & \textbf{Std.} \\ \hline
\multirow{2}{*}{\vtop{\hbox{\strut User}\hbox{\strut Satisfaction}}} 
                                & Adaptive  & 4.61 & 8.98 & 6.90 & 7.02 & 1.11 \\ \cline{2-7}
                                & NA        & 3.27 & 7.00 & 5.93 & 6.02 & 0.79 \\ \hline
\multirow{2}{*}{\vtop{\hbox{\strut User}\hbox{\strut Engagement}}} 
                                & Adaptive  & 2.26 & 4.43 & 3.41 & 3.46 & 0.61 \\ \cline{2-7}
                                & NA        & 1.90 & 4.64 & 3.18 & 3.19 & 0.57 \\ \hline
\end{tabular}
\end{table}

\subsection{Hypotheses Testing}

\begin{table*}[ht]
\centering
\caption{Fixed Effects Estimates of the LMMs.}
\label{tab:lmm}
\begin{tabular}{l|cccc|cccc}
\hline
                        & \multicolumn{3}{c}{\textbf{User Satisfaction}} & & \multicolumn{3}{c}{\textbf{User Engagement}} \\
\hline
\textbf{Predictors}                                           & Estimates & Lower CI & Upper CI & p-value & Estimates & Lower CI & Upper CI & p-value \\
\hline
(Intercept)                                & 6.50      & 5.85     & 7.14     & \textbf{$<$0.001 ***} & 3.27      & 2.90     & 3.64     & \textbf{$<$0.001 ***} \\
domain {[}trips{]}                         & -0.06     & -0.29    & 0.17     & 0.600                 & -0.07     & -0.25    & 0.12     & 0.458 \\
technique {[}Non-Adaptive{]}               & -0.97     & -1.20    & -0.74    & \textbf{$<$0.001 ***} & -0.23     & -0.41    & -0.04    & \textbf{0.016 *} \\
group {[}2{]}                              & 0.88      & -0.04    & 1.79     & 0.060                 & 0.45      & -0.06    & 0.96     & 0.085 \\
group {[}3{]}                              & 0.64      & -0.24    & 1.53     & 0.152                 & 0.17      & -0.33    & 0.66     & 0.504 \\
group {[}4{]}                              & 0.29      & -0.55    & 1.13     & 0.498                 & 0.11      & -0.36    & 0.58     & 0.631 \\
\hline
\multicolumn{9}{l}{\rule{0pt}{1em}\textit{Note: * p$<$0.05; ** p$<$0.01; *** p$<$0.001} }\\
\end{tabular}
\end{table*}

    Model assumptions were evaluated for both dependent variables (\textit{User Satisfaction} and \textit{User Engagement}). The Shapiro-Wilk test was used to confirm the normality of residuals, indicating no significant deviation from normality ($p > 0.05$). Homogeneity of variances was assessed using Levene’s test, with results ($p > 0.05$) verifying that variances were consistent across conditions. These findings confirm that the assumptions required for conducting the Linear Mixed Model analysis were satisfactorily met.
    
    Following the descriptive analysis and assumption testing, we built a Linear Mixed Model for each dependent variable. The results of the fixed effects are shown in Table~\ref{tab:lmm}. On the one hand, the analysis showed that the technique used had a significant impact on \textit{User Satisfaction}, with non-adaptive technique resulting in lower satisfaction levels (Estimate = $-0.97$, $p < 0.001$). This finding supports the hypothesis that users would express higher satisfaction with adaptive techniques compared to the non-adaptive one. Conversely, the domain variable did not exhibit a significant effect (Estimate = $-0.06$, $p = 0.602$), indicating that \textit{User satisfaction} did not significantly differ between the Courses and Trips domains. Since group was confounded with carryover and the technique*period interaction, we can conclude that none of these variables are significant. Thus, there is no carryover effect between the techniques. On the other hand, for the \textit{engagement} model, the technique used also demonstrated a significant effect, with the non-adaptive technique associated with a lower engagement level (Estimate = $-0.23$, $p = 0.016$). This result supports the hypothesis that adaptive techniques would lead to higher engagement levels compared to the non-adaptive one. As with satisfaction, the domain variable did not produce a statistically significant effect on engagement (Estimate = $-0.07$, $p = 0.458$), suggesting that engagement levels were consistent across both the Courses and Trips domains. Similarly to \textit{User Satisfaction}, there is no carryover effect between the techniques.

Table~\ref{tab:randomEffects} shows the mean random effects variance ($\alpha^{2}$) of the model and random intercept variance ($\tau_{00}$) which indicates how much subjects differ from each other.
The ICC (Intraclass Correlation Coefficient) of $76\%$ and $55\%$ indicates good and moderate reliability~\cite{terry:icc:2016} for \textit{User Satisfaction} and \textit{User Engagement} respectively, i.e. a single individual produces consistent measurements. In other words, it means more difference among individuals than within them, justifying the individual differences modeled by random effects.
Besides, the $R^{2}$ values provide further insight into the explained variance for each model. For \textit{User Satisfaction}, the Conditional $R^{2}$ was $0.828$, indicating that $82.8\%$ of the variance is explained by both fixed and random factors, with the Marginal $R^{2}$ at $0.281$, suggesting that the fixed effects alone account for $28.1\%$ of this variance. This points out to a the role of individual differences in \textit{User Satisfaction}. Similarly, for \textit{User Engagement}, the Conditional $R^{2}$ was $0.605$, showing that $60.5\%$ of the variance is explained by both fixed and random effects, while the Marginal $R^{2}$ of $0.113$ indicates that fixed factors account for only $11.3\%$ of the variance. This finding emphasizes the greater influence of individual differences on user satisfaction relative to user engagement.

\begin{table}[ht]
    \centering
    \caption{Random Effects of the LMMs.}
    \label{tab:randomEffects}
    \begin{tabular}{l|l|l}
    \hline
    \textbf{Random Effects} & \textbf{User Satisfaction} & \textbf{User Engagement} \\
    \hline
    $\alpha^{2}$            & 0.21        &  0.14     \\
    $\tau_{00}$             & 0.68 \textsubscript{user}       &  0.17 \textsubscript{user}    \\
    ICC                     & 0.76        &  0.55     \\
    N                       & 33 \textsubscript{user}          &  33 \textsubscript{user}    \\
    \hline
    Observations            & 66          & 66              \\
     Marginal $R^{2}$        & 0.281       & 0.113            \\
    Conditional $R^{2}$     & 0.828       & 0.605            \\
    \hline
    \end{tabular}
\end{table}

    In summary, these findings suggest that the adaptation strategy used in this experiment, namely Adaptive UI with individualized human feedback which create a unique RL Agent for each user, significantly enhances both \textit{User Satisfaction} and \textit{User Engagement}, independent of domain, thereby supporting the hypothesis that this adaptation technique positively impact user experience.

\subsection{Discussion}

This study provides valuable insights into individualized preference models in AUIs. Differing from our previous experiment~\cite{gasparEMSE:2024}, which found no significant differences between adaptive and non-adaptive interfaces, the current study underscores the value of adaptation strategies informed by individual user data. Unlike the earlier study, which used a generalized human feedback model, trained by an experimenter for all participants, the present study employed individualized preference models, leveraging both predictive HCI models and human feedback to train a unique RL agent for each user. This resulted in statistically significant improvements in user satisfaction and engagement, emphasizing the effectiveness of personalized adaptation strategies.


To the best of our knowledge, a broader insight is the absence of standardized methodologies to address the core challenges associated with AUIs. Key challenges include defining the AUI problem, developing adaptation strategies, evaluating adaptations, and systematically collecting and integrating human feedback. This lack of standards represents a significant gap in the field. In contrast, domains like large language models (LLMs) have greatly benefited from well-defined standards for tasks such as feedback collection, model evaluation, and adaptation, which have unlocked the full potential of human feedback~\cite{havrilla:rlhfLLM:2023,lai:okapiRLHFllm:2023}. To bring AUIs to a similar level of progress, the development of shared frameworks and benchmarks would be highly beneficial. Specifically, establishing standardized approaches for representing AUIs and gathering human feedback could enhance both the consistency and comparability of research,

\subsection{Implications for Research and Practice}

This study offers broader implications for both researchers and practitioners. For researchers, it provides a framework for capturing user feedback and incorporating it into a decision-making system that adapts the UI according to individual user preferences.
The systematic approach outlined in this paper serves as a guide for future studies aimed at refining RL-based methodologies for collecting and applying human feedback. These insights could contribute to the development of more sophisticated UI adaptation algorithms. In addition, the experimental design and materials defined in this study have been made available for replication at \url{https://figshare.com/s/4b93d59ee39853aeaa88}.

For developers, the findings underscore the practical benefits of integrating individualized preference models into adaptive systems. By employing such models, software applications can dynamically respond to user preferences, enhancing personalization and delivering a more efficient user experience. This capability is especially valuable in domains where user diversity and task complexity require highly flexible and responsive systems.

Additionally, the study provides a framework for how user feedback can shape the design of adaptive systems, offering a systematic approach to guide future UI adaptation strategies. Researchers may leverage our findings to refine research methodologies for gathering actionable feedback that informs UI adaptation algorithms, while developers can apply these insights to improve their software applications' responsiveness to user preferences, thereby delivering more personalized and efficient user experiences.

Ultimately, this study reinforces the importance of human-centered design principles in the development of adaptive systems. The improvements observed in user satisfaction and engagement highlight the transformative potential of individualized preference models, which not only advance adaptive UI strategies but also contribute to shaping the broader landscape of user interaction design. These findings inspire further exploration into personalized adaptation techniques.

\section{Threats to Validity}

This section addresses potential threats to the experiment's validity and how we attempted to mitigate them~\cite{Wohlin:2012}, categorized as internal, conclusion, construct, and external.


\textit{Internal validity threats}, which concern influences on the independent variables, primarily included learning effects, the Hawthorne effect~\cite{adair1984hawthorne}, and maturation effects. To mitigate these, we ensured all participants interacted with both \textit{Adaptive} and \textit{non-adaptive} UIs, and a balanced crossover design was employed to control for sequence effects. Participants were instructed to behave naturally to minimize the Hawthorne effect, and breaks were incorporated to reduce maturation effects.


\textit{Conclusion validity threats}, related to the accuracy of inferences drawn from the experiment, were addressed through robust experimental practices. These included recruiting an adequate sample size to ensure sufficient statistical power and selecting appropriate statistical tests based on data characteristics and assumptions~\cite{Maxwell2002Applied, vegas:2015}.


\textit{Construct validity threats}, concerning the ability to generalize results to the supporting theory, focused on the validity and reliability of questionnaires. Strong reliability was demonstrated with Cronbach’s alpha for QUIS ($0.97$) and UES ($0.90$), indicating excellent internal consistency. Item-total correlations were also analyzed, showing acceptable values.


\textit{External validity threats}, related to the generalizability of findings to broader contexts, were limited by the focus on specific tasks and controlled conditions, which may not fully replicate the complexity of real-world interactions. Mitigation involved designing realistic and relevant tasks for the e-learning and trip planning domains. However, generalizing to more diverse contexts, such as different platforms or environments, remains a limitation. Future work should consider a wider range of contextual factors, such as platform and environmental variability, to enhance generalizability.


\section{Conclusions and Further Work}

We extended a RL-based framework for AUIs by incorporating individualized human feedback into the reward modeling process of RL agents, enabling the creation of personalized adaptation strategies. We evaluated this extended framework through an empirical study with human participants, who initially provided feedback on various UI adaptations. This feedback was then used to train personalized RL agents for each participant. Participants interacted with two systems (a trip planning platform and an e-learning platform) each employing two techniques: a RL-based AUI utilizing the personalized agents and a non-adaptive interface. The findings suggest that incorporating individualized models derived from human feedback significantly enhances both user satisfaction and user engagement.

Further work should explore the scalability of these individualized models across more domains to improve generalizability and investigate real-time feedback integration. Additionally, investigating the feasibility of training a single preference model to develop cluster-based models using shared user profile characteristics would be valuable. We also plan experiments to enhance the transparency and explainability of our RL-based framework for AUIs.
Moreover, unlike other domains with established standards for ML model development and human feedback collection, AUIs currently lack such standardized approaches. Establishing a standardized approach could create a baseline for more scalable and consistent solutions, making AI-driven UI adaptation more feasible and improving system efficiency and effectiveness across applications.

\begin{acks}
This work was funded by the GVA under the AKILA project (CIAICO/ 2021/303) and by the AEI under the UCI-Adapt project (PID2022-140106NB-I00). D. Gaspar-Figueiredo is funded by the GVA (ACIF/ 2021/172), which is cofunded by the EU through the ESF.
\end{acks}

\bibliographystyle{ACM-Reference-Format}
\bibliography{Arxiv/main}


\end{document}